\documentclass{elsart}

\usepackage{epsfig}
\usepackage{amssymb}

\begin{document}

\begin{frontmatter}

\title{A Method to Search for Local Sources of Short Duration Bursts
       of Superhigh Energy Gamma Rays}

\author{E.N.Alexeyev, D.D.Djappuev, A.U.Kudjaev, S.Kh.Ozrokov}

\address{Institute for  Nuclear Research, Russian Academy of Sciences,
60th October Anniversary Avenue, 7a, Moscow 117231, Russia}

\begin{abstract}
A method of a search for  local sources of superhigh energy gamma rays
is  described  in  the  paper.It  is shown  that  the  method  is more
effective then the usually used  method  extracting  excess from total
intensity if gamma ray  burst  durations are short.Using the suggested
method,the   information   detected   with  the  Baksan   installation
``Carpet'' during  1992-1996  years  was  analyzed.An  excess of event
numbers  was  found at  the  confidence level  of  6.5$\sigma$ in  the
direction to Mrk 501.
\end{abstract}

\end{frontmatter}

\section{Introduction}

The considerable progress began to show in astrophysics of high energy
gamma  rays  at  the  present  time.  Using the terrestrial  Cherenkov
telescopes, gamma  rays from cores of  active galaxies Mrn  501 and
Mrn 421,  which are located at the  distances of  about 150 Mpc,  were
detected  \cite{c1,c2}.  The signal  from  the  Crab  nebula  detected
formerly by  the same telescopes  is presently used as the ``standard''
candle. That success in high energy gamma ray astronomy  is mainly due
to Imaging Atmospheric Cherenkov Technique  originally proposed in 1977
\cite{c3}. Meanwhile,  there are no  reliable data on the detection of
superhigh energy gamma rays from local sources using 
Extensive Air  Shower (EAS) arrays except the Tibet detector \cite{c4}
which detected gamma rays from Mrn 501 in the similar energy region
(about 10  TeV). Revealing local sources  which emit  gamma rays in
superhigh energy region (above 100 TeV) would allow us to make a step
to understanding the nature of superhigh energy particle acceleration.

In the  paper  a method of a detection of gamma rays with energy of $E
\ge 10^{14} eV$  is  suggested. The method has  a  high sensitivity to
local  gamma  ray  sources in the  case  of  the  radiation from these
sources to be discrete one and  to have duration of about 1 second for
any burst, if such sources really exist in our universe.

\section{The Method of the Search for Local Sources}

The main  point of the method consists  in the  selection of pair
events every one of which arrived in the detector from the  same  direction 
and a time  interval $\Delta{t}$ between  them is short. It is not difficult to
show that  for some groups of superhigh energy  gamma ray sources with
short burst duration, the suggested method is more  effective than the
usually used one which looks for exceeding over  background in summary
intensity.

Assume $m_\gamma$ to be a number of events which arrive at a  detector
from any source within an  angle cell of sky sphere with  dimensions 
of $\Delta\alpha$ and $\Delta\delta$ during time  interval $\tau$. Let
 $N$ to be a number of such bursts within an observation time  $T$. Then
the efficiency  of the search  using summary intensities will be equal
to following
\begin{eqnarray}\label{e01}
E_1 = \frac{m_\gamma \cdot N}{n_f \cdot T},
\end{eqnarray}
where $n_f$  is a background counting rate within an  angle cell in the
direction to a supposed source.

The search  efficiency for the above  mentioned pair events defined by the
similar way is the next
\begin{eqnarray}\label{e02}
E_2=\frac{m_\gamma^2\cdot N}{\tau\cdot n_f^2\cdot T}.
\end{eqnarray}

Dividing the  equation  (\ref{e02})  by  the  equation (\ref{e01}), we
obtain the ratio  of  those two kinds of  efficiencies
\begin{eqnarray}\label{e03}
\frac{E_2}{E_1} = \frac{m_\gamma}{\tau \cdot n_f}.
\end{eqnarray}

The average counting  rate  of events  at  the detector ``Carpet''  is
equal  to   $n_f\sim   10^{-4}$sec$^{-1}$   within   angle   cell   of
$3^{o}\times3^{o}$. So, as it is seen from Eq. (\ref{e03}), the search
efficiency of local sources using pair events will be better  than that
in the case using  summary intensities if  mean number of gamma rays arriving
at the ``Carpet'' during  the  burst is $m_\gamma > 10^{-4}$
and the burst time duration  is $\tau <$ 1 sec.

Here  it is  necessary to  note  that for the  given method  of a local
source search an  existence of a periodicity of a source radiation is not
the   necessary   condition.  Nevertheless,  in  the  presence  of   a
periodicity it is possible to evoke  an  additional  information  and,
probably, to identify the source.

The suggested pair  event  method of the search  for  local sources of
superhigh energy gamma rays  was  tested with the information detected
by the installation ``Capret'' in the period of 1992-1996 years.

\section{The Baksan installation ``Carpet''}

The  installation   ``Carpet''   was  described  in  detail  elsewhere
\cite{c5}. Here  we only remind of its main  parameters. It is located
at  the Baksan  Neutrino  Observatory in the  mountains  of the  North
Caucasus.

The  central  installation part  with  the  total  area  of  196 m$^2$
consists of 400 standard liquid  scintillator  detectors  every one of
which  has the area of 0.49 m$^2$.

There are  4 carrying-out  points at a distance of  30 meters from the
installation center and  2  carrying-out points  at  a distance of  40
meters.  Every  carrying-out  point  consists  of  the  same  standard
scintillator detectors, number of which  is equal to 18. This means that
the every point area is equal to 9 m$^2$.

The muon detector with the area  of 175 m$^2$ is located at a distance
of 48 meters from the installation center. Unfortunately,  it began to
work  later  and  its  data  could  not  be  used  in   the  analysis.
Eight-multiple coincidences of  four quarters of the central part with
four nearest (located at a distance of 30  meters) carrying-out points
are the installation trigger impulses. In this case  the counting rate
of showers is equal to $\sim$1 sec$^{-1}$ and the EAS threshold energy
is about $E\ge 10^{14} eV$. The angle accuracy is
not worse than $1.5^{o}$.

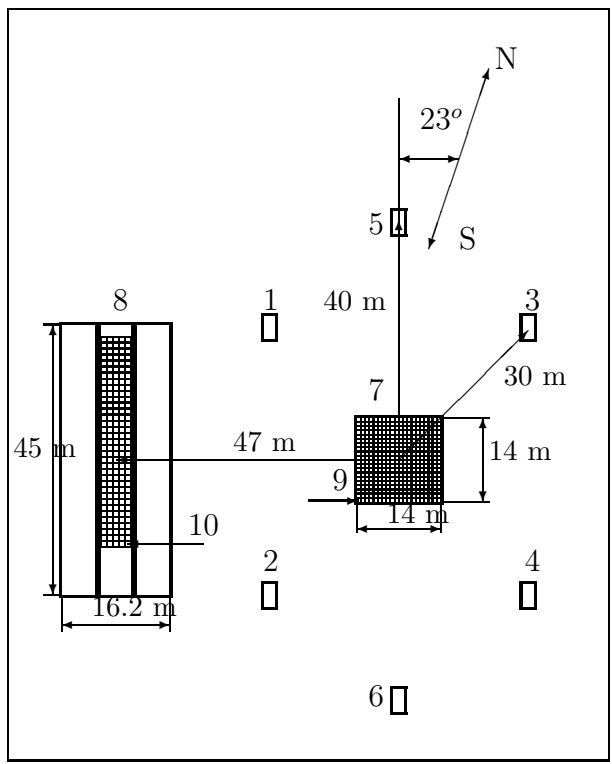
\begin{figure}
\begin{center}
\setlength{\unitlength}{.4mm}
\begin{picture}(180,250)(15,0)

\normalsize
\thicklines
\multiput(28,55)(12,0){3}{\framebox(12,90)}
\multiput(95,51)(86,0){2}{\framebox(4,8)}
\multiput(95,140)(86,0){2}{\framebox(4,8)}
\multiput(138,16)(0,159){2}{\framebox(4,8)}
\put(126,86){\framebox(28,28)}

\thinlines
\multiput(127.4,86)(1.4,0){19}{\line(0,1){28}}
\multiput(126,87.4)(0,1.4){19}{\line(1,0){28}}

\multiput(41,141)(0,-2.){36}{\line(1,0){10}}
\multiput(41,141.)(2.,0){6}{\line(0,-1){70}}

\put(140,180){\line(0,1){40}}
\put(160,200){\vector(1,3){10}}
\put(160,200){\vector(-1,-3){10}}
\put(172,230){N}
\put(160,170){S}
\put(147,210){$23^o$}
\put(147,200){\vector(1,0){13}}
\put(147,200){\vector(-1,0){7}}

\footnotesize
\multiput(28,55)(0,90){2}{\line(-1,0){6}}
\put(25,55){\vector(0,1){90}}
\put(25,70){\vector(0,-1){15}}
\put(12,101){45 m}

\multiput(28,55)(36,0){2}{\line(0,-1){12}}
\put(28,45){\vector(1,0){36}}
\put(43,45){\vector(-1,0){15}}
\put(38,48){16.2 m}

\multiput(126,86)(28,0){2}{\line(0,-1){10}}
\put(140,77){\vector(1,0){14}}
\put(140,77){\vector(-1,0){14}}
\put(136,79){14 m}

\multiput(154,86)(0,28){2}{\line(1,0){16}}
\put(168,100){\vector(0,-1){14}}
\put(168,100){\vector(0,1){14}}
\put(170,100){14 m}

\put(160,120){\vector(1,1){23}}
\put(160,120){\line(-1,-1){20}}
\put(175,125){30 m}

\put(46.25,100){\line(1,0){94}}
\put(62,100){\vector(-1,0){15.75}}
\put(85,103){47 m}

\put(140,140){\vector(0,1){40}}
\put(140,140){\line(0,-1){40}}
\put(115,150){40 m}

\normalsize
\put(95,150){1}
\put(182,150){3}
\put(95,63){2}
\put(182,63){4}
\put(130,18){6}
\put(130,175){5}
\put(130,120){7}
\put(45,150){8}
\put(110,86.5){\vector(1,0){16.5}}
\put(118,90){9}
\put(75,72){\vector(-1,0){25}}
\put(70,75){10}

\multiput(10,0)(0,250){2}{\line(1,0){200}}
\multiput(10,0)(200,0){2}{\line(0,1){250}}

\end{picture}
\end{center}
\caption{The  scheme  of  the  installation of ``Carpet'',  the  graph
points are: ``1-6''  mark carrying-out points, ``7'' shows the central
part, ``8'' is the muon detector, ``9'' shows  the liquid scintillator
detector, ``10'' is the plastic scintillator detector.}
\end{figure}

\section{The Event Analysis}

In the equatorial  system of coordinates $\alpha$, $\delta$ , 
an optimum size of  an  angle  cell  was  chosen  in  order to get the
maximum ratio of an expected signal to background one. The cell size is to
content with  a condition  that 80\% of source events  will hit in it.
This means  that at the angle accuracy  of $\Delta  R(\theta)= 1.5^{o}$ the cell
size required is to be equal to $3^{o}\times3^{o}$.

The total number of  events  detected with the installation ``Carpet''
in the  period of 1992-1996 years made up  $\sim 10^{8}$ showers. From
this field of events,  event  pairs contenting the following selection
criteria were chosen:
\begin{enumerate}
\item an angle distance between pair events is to be $d \le 5^{o}$;
\item time interval between pair events  is  to  be  $\Delta{t}\le$150
msec.
\end{enumerate}

\section{Results}

The suggested method of ``pair  events''  was used in the analysis  of
data detected from different directions to  galactic  objects  and  to
extragalactic ones, namely  Mrn  501 and Mrn 421.  The  results of the
observation of pair events from the last two sources are  presented in
this paper.

The selected  pair events were accommodated  in sky sphere  cells with
the dimensions of $3^{o}\times3^{o}$ under condition  that both events
of  the  pair  are  containing  inside  one cell.  In  the  equatorial
coordinates  ($\alpha$,  $\delta$)  a  layer  on  the  sky  sphere was
considered, where
\begin{eqnarray}
38^{o} \le \delta \le 41^{o}. \nonumber
\end{eqnarray}
The   layer   consisted   of   120  cells  with  the   dimensions   of
$3^{o}\times3^{o}$.

The cell with the direction to Mrn 501 keeps the position number 84 in
this layer. Using the average number of event pairs in the layer cell,
the excess  $q$ within the given cell was  determined by the following
formula in the units of standard deviation $\sigma$
\begin{eqnarray}
q=\frac{(n_{84}-\overline{n})}{\sigma}, \nonumber
\end{eqnarray}
where $n_{84}$  is the number of event  pairs within  the cell in  the
direction to Mrk 501, \\
$\overline{n}$ is the average number of pairs in the layer cells, \\
$\sigma$ is the  mean  square root deviation of  pair  number over the
layer.

The analysis was  performed for every  year separately in  the  period
1992-1996 years. The results obtained are shown in Table 1.

\begin{table}[htb]
\caption{}
\label{table:1}
\begin{center}
\begin{tabular}{|c|c|c|c|c|c|}
\hline
year & number of & average number       &
                                      & observation & Source \\
     & pairs in  & of pairs in the cell &
 $\frac{n_{84}-\overline{n}}{\sigma}$ & time        & name   \\
     & the cell  & over the layer       &
                                      & (days)      &        \\ \hline
1992 & 3  & 4.7 & -0.8 & 362 & Màª501 \\ \cline{1-5}
1993 & 14 & 2.8 & 6.5  & 304 &        \\ \cline{1-5}
1994 & 0  & 2.8 & -1.6 & 207 &        \\ \cline{1-5}
1995 & 4  & 4.6 & -0.3 & 344 &        \\ \cline{1-5}
1996 & 4  & 2.7 &  0.9 & 205 &        \\ \hline
\end{tabular}\\[2pt]
\end{center}
\end{table}

As it is seen from the Table 1, there  is  a  considerable  excess  of
event pairs over background ones at a confidence level of 6.5 $\sigma$
in the direction to Mrn 501 in 1993 year.

Then another layer with coordinates of  $36.5^{o}\le\delta\le39.5^{o}$
was  considered in  the  similar way. The  cell  keeping the  position
number 56 corresponds to the direction  to the object Mrn 421. In this
case the excess of the expected events relatively to average number of
events was not found in 1993 year (2 pair events relatively background
ones  of 2.9).

The temporal distributions of pair events detected  during 1992 and
1993 years from the direction to the object Mrn 501 are shown in Fig 2(a,b).
It is very interesting to remark that in 1993 year three pair events were detected on 324th day.

Figure 3 shows the  distribution of the cell numbers as a  function of
the number of pair events  within  cells detected from the sky  sphere
layer containing the object  Mrn 501 in 1993 year. The excess  of pair
events which is clearly seen in the last figure has a confidence level
of 6.5  $\sigma$.

Fig.4 shows the  distribution of single  events within cells  for the  same
layer containing Mrn  501  detected during  1993  year. The number  of
single events (11762) in  the direction to Mrn 501 is shown  by the pointer.
Mean  number of  events  over the layer is 11817. As one can see from the
Fig.4  there is no  any excess of single events  in  the direction  to Mrn  501.

If the pair event excess in the direction to Mrn  501  was caused by
real gamma rays,  it  is necessary to assume  that  the radiation from
this source was increased in 1993 year. The similar phenomenon, namely
the increase of  the  Mrn 501 radiation intensity  in  the high energy
region,  was  observed   in  1997  year.  Unfortunately,  we  have  no
possibility  to test  the  availability of the  signals  from Mrn  501
during 1997 year because  the  installation ``Carpet'' did not operate
in that time.

\begin{figure}
\epsfxsize=.9\hsize
\centerline{\epsfbox{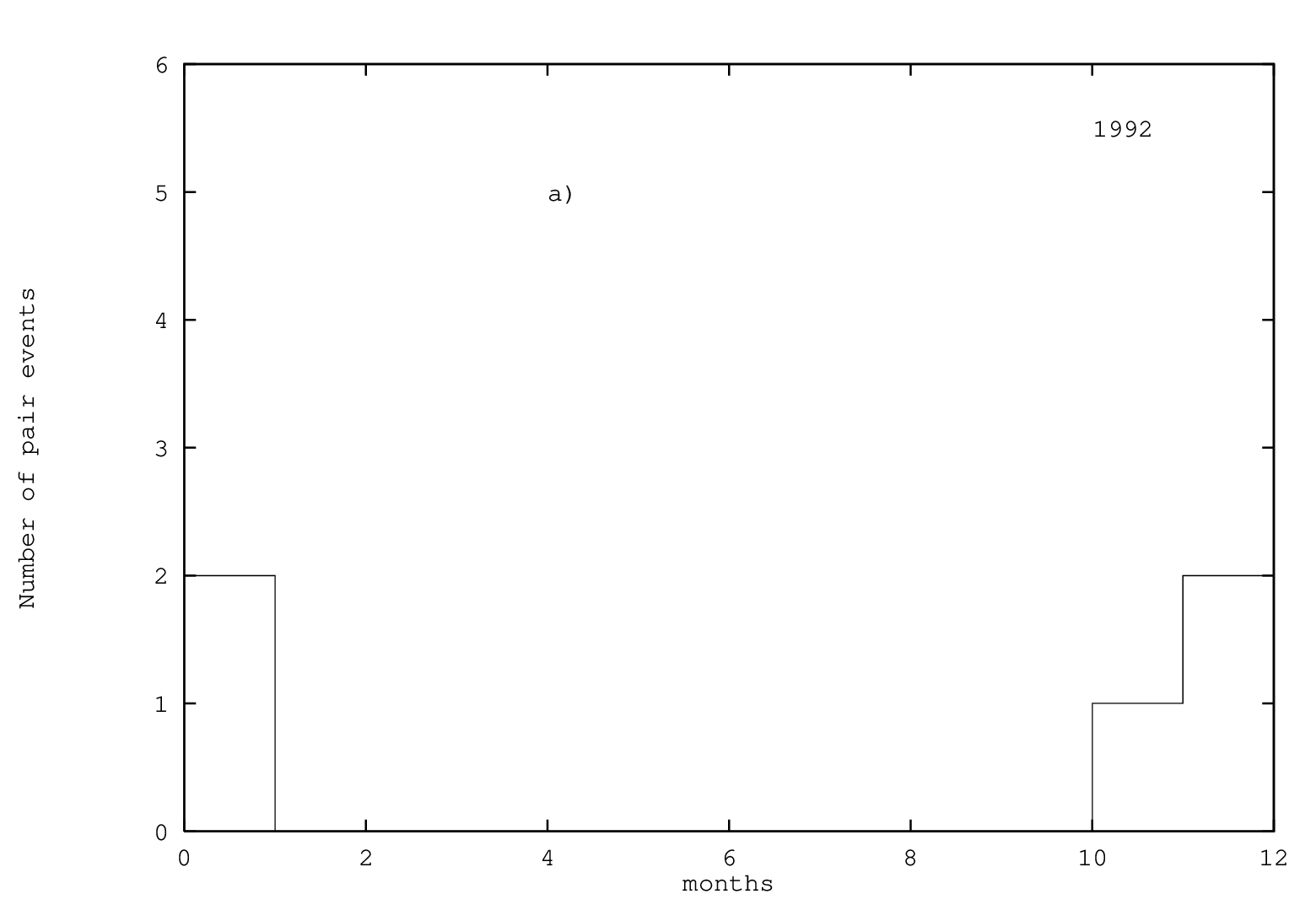}}
\epsfxsize=.9\hsize
\centerline{\epsfbox{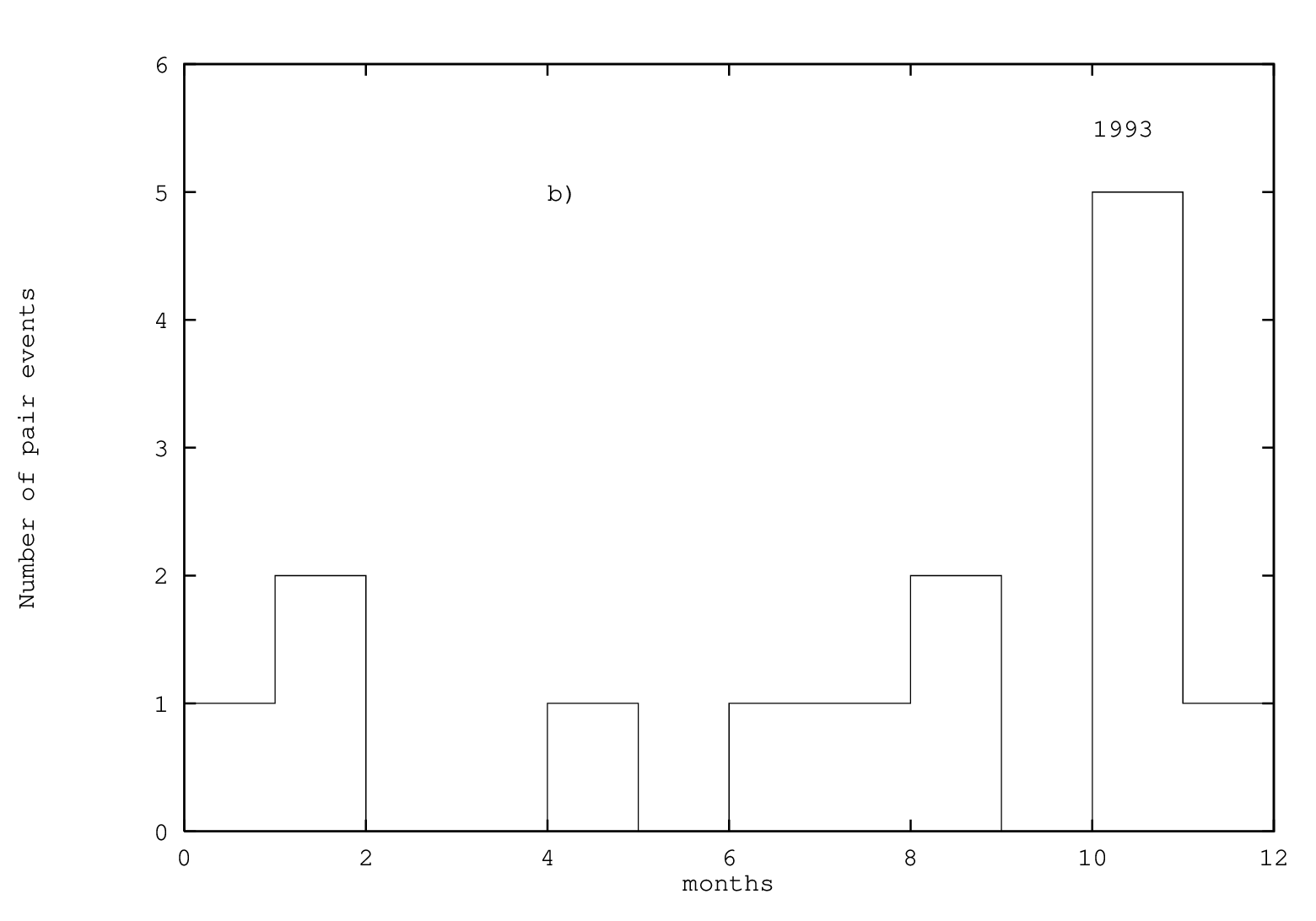}}
\caption{The temporal distribution of pair events detected during 1992
and 1993 years from the direction to Mrk 501.}
\end{figure}

\begin{figure}
\epsfxsize=.9\hsize
\centerline{\epsfbox{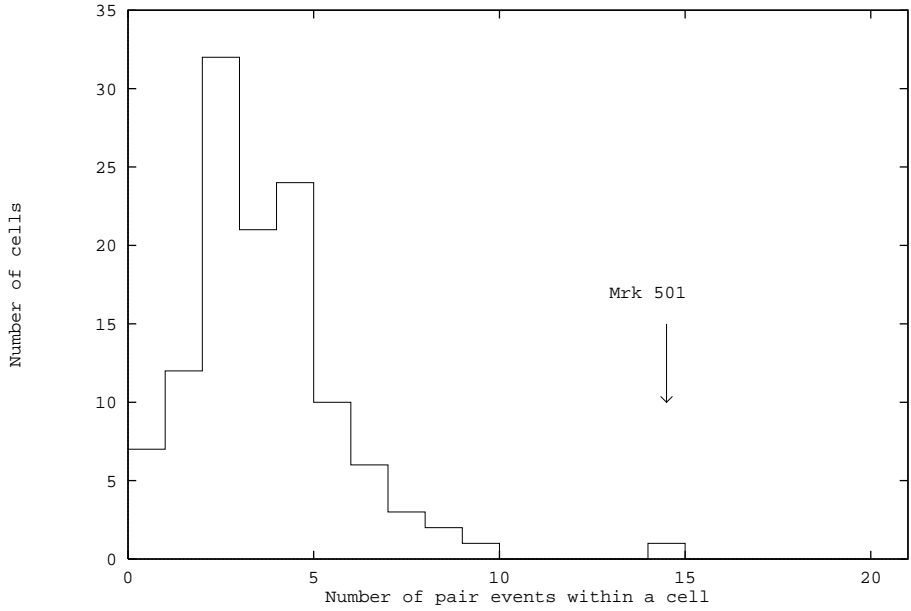}}
\caption{The distribution of  the cell numbers  as a function  of  the
number  of  pair events within the  cells  detected in the sky  sphere
layer containing the source Mrk 501.}
\end{figure}

\begin{figure}
\epsfxsize=.9\hsize
\centerline{\epsfbox{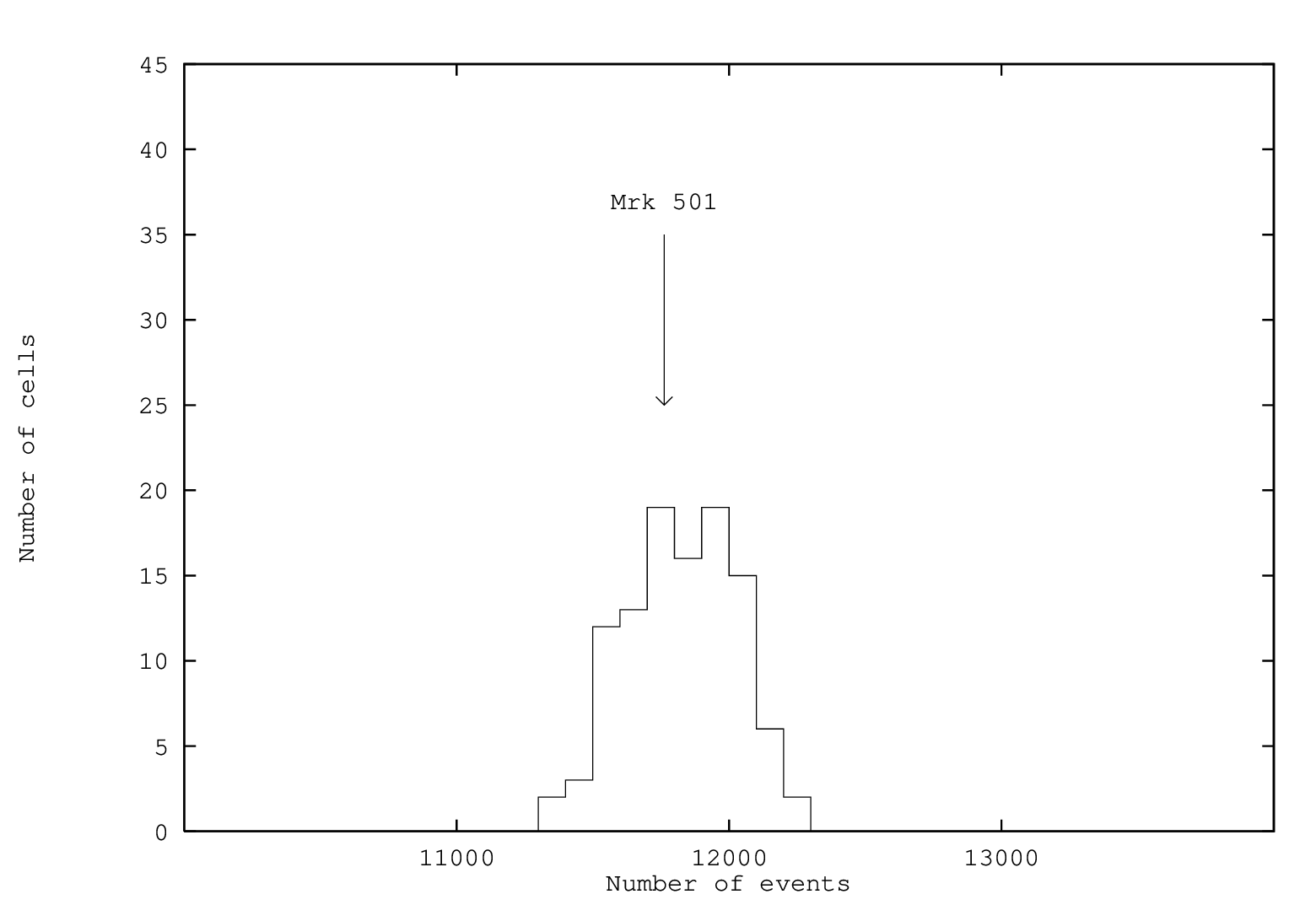}}
\caption{The  distribution  of single events detected  during  1993
year in  layer which contains  the Mrn  501 cell. 
The number of events in the Mrn 501 cell  is shown with the pointer.}
\end{figure}

\section{Conclusion}

Using the data detected by the Baksan installation ``Carpet'' with the
Extensive  Air  Showers  energy  threshold of $E\geq 10^{14}  eV$,  the
method of the search for  local  sources by analyzing pairs of  events
from different galactical and extragalactical sources  was tested. The
signals from some galactical sources were found at  a confidence level
of about  4 $\sigma$. So, in order  to confirm  the reality of  these
signals, the analysis of  joint  data from several detectors performed
with the same method is required.

The  excess of pair  events  is  found  within sky  sphere  cell  with
dimensions of $3^{o}\times 3^{o}$ from the direction to the object Mrn
501. The  excess confidence level  is equal to 6.5$\sigma$.

Taking into account the high value of the  ``Carpet'' energy threshold
(100 TeV)  and the large distance to  the source,  it is difficult  to
explain  the  event excess from the  direction  to the source Mrn  501
observed in 1993 by the detection of Extensive Air Showers produced by
its gamma rays because gamma rays will interact with infrared radiation 
during its way from Mrn 501
\cite{c6}. One of possibilities to explain the effect can be a location
of some  galactical gamma  ray source at the direction  to Mrn 501. In
any case, it  would be extremely  important to test  this  information
with other installations.

And  in the conclusion  it  is  necessary  to  note  once more  that an
availability of a signal from some source detected  by the presented
method of  pair events will be  the evidence  of an unusual  nature of  the
source emission,  namely, of the  short duration (about one second) of
particle bursts. If the particles which were detected in this experiment
were really emitted by Mrn 501, it means that the time scale of the
acceleration of superhigh energy particles in the AGN is $\tau < $ 1 sec.

\end{document}